----------------------------- Start of body part 2

\documentstyle[12pt]{article}

\begin{document}
\pagestyle{empty}
\begin{titlepage}
\title{A brief history of hidden quantum symmetries \\ in \\
Conformal Field Theories\thanks{Lecture presented in the XXI DGMTP-Conference
(Tianjin ,1992) and in the NATO-Seminar (Salamanca,1992).} }

\author { C\'esar G\'omez and Germ\'an Sierra \\
Instituto de Matem\'aticas y F\'{\i}sica Fundamental, CSIC \\
Serrano 123.  Madrid, SPAIN}
\date{}
\maketitle

\begin{abstract}
We review briefly a stream of ideas concerning the role of quantum
groups as hidden symmetries in Conformal Field Theories, paying
particular attention to the field theoretical representation of
quantum groups based on Coulomb gas methods. An extensive bibliography
is also included.

\end{abstract}

\end{titlepage}
\pagestyle{plain}

\subsection*{Integrability, Yang-Baxter equation and Quantum Groups}

\noindent
In the last few years it has become clear the close relationship
between conformal field theories (CFT's), specially the rational
theories , and the
theory of quantum groups.From a more general point of view
this is just one aspect of the connection between integrable
systems and quantum groups. In fact if one recalls the history,
quantum groups originated from the study of integrable models
and more precisely from the quantum inverse scattering approach
to integrability. The basic relation:

$$
R(u - v)\;  T_1(u)\; T_2(v) = T_2(v)\; T_1(u)\; R(u -v)
\eqno(1)
$$

\noindent
which
is at the core of this approach may also be taken as a starting
point in the definition of quantum groups [FRT].

In statistical mechanics the operator $T(u)$ is nothing but the monodromy
matrix which depends on the
spectral parameter $u$ and whose trace gives the transfer matrix of
a vertex  model. Meanwhile
the matrix $R(u)$ stands for the Boltzmann weights of this vertex model.
The $RTT$ equation guarantees the commutativity of a one
parameter family of transfer matrices:

$$
[\; tr \;T(u) ,\; tr\; T(v)\; ] = 0
\eqno(2)
$$

This equation is equivalent to the existence of an infinite
number of conserved quantities when taking the thermodynamic limit.
In these setting the Yang-Baxter equation:

$$
R_{1 2}(u) R_{1 3}(u + v) R_{2 3}(v) =
R_{2 3}(v) R_{1 3}(u + v) R_{1 2}(u)
\eqno(3)
$$

\noindent
which underlies the integrability
of the model appears as the associativity condition of the "quantum
algebra" (1).

\subsection*{RCFT's, exchange algebras and the braid group
: the polynomial equations}

\noindent
Let us turn now to conformal field theories. There, it was soon realized
that the braiding matrices of the conformal blocks of a RCFT
also satisfy a Yang-Baxter like equation. In fact these matrices
provide a representation of the braid group in the space of conformal
blocks. The braid group and the exchange
algebras appeared suddenly to play an important role
in CFT's and more generally in 2D quantum field theories
[TK,Ko,Fr,FRS,Rh,MS].
However there are some structural differences with regard
to the vertex models in statistical mechanics.
In CFT the braiding matrices do not depend on any
spectral parameter while in statistical mechanics the Boltzmann
weights certainly do  depend on it and only when taking the
limits $u \rightarrow \pm \infty $ one gets non trivial solutions
of the Yang-Baxter equation without spectral parameter.
In addition ,the
Boltzmann weights of a
vertex model $R_{i j}^{k l}(u)$
depend on four labels  while
the braiding matrices
$B_{p p'} \left[ \begin{array}{cc}
k & l \\
i & j
\end{array}
\right]$
depend on six labels which
are the primary fields of a RCFT and consequently
are subject to certain constraints given by the fusion rules.
These later conditions are equivalent to the restrictions
on the heights of a restricted solid on solid model ( RSOS) where
the Coxeter diagram is playing the role of the fusion rules in RCFT.

Having in mind the $RTT$ equation of integrable models as well
as the braiding properties of the conformal blocks it was
proposed in reference [AGS1] that quantum groups should be present
in RCFT's as "hidden symmetries" underlying and explaining the origin
of the polynomial equations of Moore and Seiberg. The motivation of these
authors in writing the polynomial equations was to give an axiomatic
definition of RCFT's which would, among another things, give a rigorous
proof of the Verlinde conjecture concerning the diagonalization of
the fusion rules $N_{i j}^k$ by means of the modular matrix
$S_{i j}$ which acts on the Virasoro characters [V].

Reference [AGS1] was in this way a first attempt to lay down a "quantum
group" interpretation of the duality properties of RCFT's, which are
encoded in the polynomial equations.
Indeed one of these
equations is nothing but the Yang-Baxter equation for the braiding
matrices of the conformal blocks mentioned above.
A second attempt in the same direction was taken by Moore and
Reshetikhin in reference [MR].

A first clue of the problem is suggested by the classical limit
of a WZW model based on a group $G$ which is obtained when taking
the level $k$ going to infinity. This model has an infinite
number of primary fields labelled  by their spin $j = 0,1/2,1,\dots$
whose braiding and fusion properties can be computed entirely from
the representation theory of the classical group $G$ [MS]. Thus for example
the fusion matrices $F$ of this classical WZW model become the
6-j symbols of the group $G$ , while the fusion rules are nothing
but the usual Clebsch-Gordan decomposition of the tensor product
of irreps of G. More generally, what one calls the
duality data of a RCFT, namely: fusion rules, braiding matrices,
fusion matrices, modular matrices, etc. are computable in the
classical limit using ordinary group theory. Of course not all
the RCFT's known admit such a classical
limit where all the conformal weights
of the primary fields vanish and consequently
the braiding properties become almost trivial.
One example of this is provided by the $c < 1$ theories.
However what this limit
suggested was that some "deformation" of the "classical" group theory
could in principle explain the whole duality data of genuine quantum
RCFT's.

\subsection*{Quantum group phenomenology}

\noindent
At that time quantum groups have already taken off from the land
of integrable systems.
The general notion of quantum groups were developed
by Drinfeld
and Jimbo in references [Dr1,Ji].
The first non classical
example of this definition was a q-deformation of the Lie algebra
$Sl(2)$ which was found earlier by Kulish and Reshetikhin [KuR].
Drinfeld and Jimbo generalized this quantum
deformation to include all the classical Lie groups affine or
not.
Later on in a series of papers Kirillov and Reshetikhin
[KR,R] constructed the representation theory of $U_q (Sl(2))$ and others
groups for a generic value of the deformation parameter $q$.

After
these works it was rather clear that quantum groups were the correct
"deformation" of classical groups needed to explain the duality properties
of RCFT's. In reference [AGS2] it was considered the case of the
WZW model based on the Kac-Moody algebra $\widehat{SU(2)}$ at level k,
and it was shown that the braiding matrices , computed by Tsuchiya and Kanie
[TK] solving the Knizhnik-Zamolodchikov equation for
the conformal blocks [KZ], were in fact given by the representation
theory of $U_q (Sl(2))$ with $q = e^{ \pi i/ k + 2}$. This work
was later generalized to other groups, where the
previous relation between the quantum deformation parameter
and the data of the WZW model reads:

$$
q = e^{ i \pi/ k + g}
\eqno(4)
$$

\noindent
with k the level and g the dual Coxeter number of $G$.
It was also possible to understand
the modular properties and in particular the Verlinde theorem using
exclusively quantum-group tools [AGS3]. In what
concerns modular properties, these lead to
the concept of ribbon Hopf algebras
[RT].

Similar results were obtained for the minimal models of type (p,p'),
where the braiding matrices essentially factorize into the product of
two quantum 6-j symbols with deformation parameters: $q = e^{i
\pi p/ p'}$ and $q' = e^{i \pi p'/p}$ [AGS1, FFK].
Also some orbifold models where shown to have a quantum group structure
[DPR] given by quasi-Hopf algebras [Dr2].

The previous examples showed that the deformation parameters coming
from RCFT's were given by roots of unity.
The representation theory of quantum groups for generic values
of q is essentially the same as the classical one [Ro].However
quantum groups  at roots of unity were shown to display
unexpected and interesting properties [L,RA].
The trouble with roots of unity is that many of the
formulas in the representation theory for q generic break down or
become ambiguous. A solution to this problem, which may look a bit
ad hoc, is to restrict the representations to the good ones,
so that one has a well defined highest weight
theory, and throwing away the
bad representations [L,PS,AGS2-3].
For example for $U_q (Sl(2))$ the good representations
,also called regular or of type II, have spin j between zero and
k/2, and are in one to one correspondence with the integrable irreps
of the WZW model $\widehat{SU(2)_k}$.
This truncation was also seen to occur in the representation
theory of the Hecke algebra $H_N(q)$ which is the centralizer of
the tensor product of the spin 1/2 representation of $U_q(Sl(2))$
[Wz,AGS2]. The relation between fusion rules of RCFT's and quantum
groups have been further studied in [GP,FGP,FD,T].

These efforts of trying to
match bits of data from different
pieces of mathematics and
physics were in a way "phenomenological", because it was
missing a neat understanding of the close relationship
between RCFT's and quantum groups. Further research has
deepen and broaden the knowledge of the topic, which cannot yet considered
as being closed.

One astonishing aspect of the connection between RCFT's and quantum
groups is that on one side one is dealing with an
infinite dimensional symmetry i.e.
the Virasoro algebra , Kac-Moody algebra ,etc. and on the other the
symmetries, although a bit peculiar, are finite dimensional. How could
it be possible that some finite dimensional structure would know so much about
an infinite dimensional one?. Making a very rough analogy we
could say that given a function of the form:
$ f( z ) = z^{\alpha} \; \sum a_n z^n$ the information that quantum
groups are telling us is the integer part of $\alpha$, which is what
determines the monodromy of the function ,i.e. $f(e^{2 \pi i} z) =
e^{2 \pi i \alpha} f( z) $. In RCFT's the
role of the function $f(z)$ is
played by the conformal blocks while the phase factors
$e^{2 \pi i \alpha}$ or $e^{\pi i \alpha}$ become the
monodromy or braiding matrices respectively.
To complete this analogy we could
say that the chiral algebra
(i.e. Virasoro, Kac-Moody ,etc) would imply a differential equation
for the function $f(z)$ , equation that would contain in a disguised
way the monodromy of their solutions.

In this context
the connection  between RCFT's  and quantum groups is the relation between
a differential equation and the monodromy or braiding properties
of their solutions. The later problem is deeply connected to the
classic and well known Riemann-Hilbert problem , which consist
in the characterization of a differential equation and
the set of the solutions from the knowledge of their monodromy properties.
What the polynomial equations establish
are the consistency conditions to
be satisfied by the braiding and fusion matrices
associated
to a set of differential equations which are in principle unknown.
The "reconstruction program",
advocated by Moore and Seiberg , which tries to determine from
every solution to the polynomial equations a set of differential
equations whose solutions give the conformal blocks of the theory,
is  nothing else but a Riemann-Hilbert problem.
This program has only been sketched but, as
one can imagine, the task of finding  a solution
to the polynomial equations seems already quite formidable
without the help of some guiding principle.

An important thing that we must not forget is the meaning of the
differential equations satisfied by the conformal blocks. They guarantee
the decoupling of null vectors of the degenerate primary fields.
The very existence of these fields is what makes the theory rational
(finite number of primary fields ) and solvable.
We may expect that quantum groups
would also have a lot to say about the decoupling of null
vectors of the corresponding primary fields.

\subsection*{Vertex formulation of CFT's}

\noindent
{}From the previous discussion it seemed natural that the ordinary
formulation of RCFT's as the representation of a chiral algebra
(or rather the product of a holomorphic
$ {\cal{A}}_L$ times an antiholomorphic
chiral algebra ${\cal{A}}_R$ ) should be enlarged in order
to accommodate the quantum group symmetry, which otherwise
remains hidden or invisible. A signal of this hidden nature was that
objects like 3j-symbols or the $R$-matrix itself did not show up in
RCFT's while for example 6j-symbols did. In general the IRF
content of quantum groups was clearly present in RCFT's. This
suggested that what was needed, to uncover the quantum group structure
of RCFT's, was a "vertex formulation" of conformal field theories.
In statistical mechanics it is well known that some models admit
both kinds of formulations, either as vertex or as IRF models, in
which case there exist a vertex-IRF map which establishes the
correspondences [Ba].It happens also, when going from the vertex to
the IRF formulation, that one misses along the way the degrees of
freedom on which the quantum group is acting.These kind of
arguments were
also used in [Wi] in an attempt to derive a quantum group structure
from the 3D Chern-Simon theory.

In group theoretical
language the vertex formulation is like working with tensor products
of representations of a group, while the IRF formulation is like
performing the tensor product of these irreps and keeping only the
spaces on which the centralizer of the group is acting:

$$
V_{1/2} \otimes \cdots^N \cdots \otimes V_{1/2} = \bigoplus_j ( W^N_j
\otimes V_j )
\eqno(5)
$$

In the vertex formulation the group or q-group is acting on each
individual vector space, $V_{1/2}$ in the example above, by means
of the comultiplication. The vector space $W^N_j$ is formed by the
invariant tensors of the q-group and serves as the
representation space of the centralizer $C^N_{1/2}$ which
is defined as the set of endomorphisms of $V^{\otimes N}_{1/2}$ that
commute with the action of the group. The dimension of $W^N_j$ equals
the multiplicity of the spin j irrep into the tensor product
$V^{\otimes N}_{1/2}$. In RCFT's $W^N_j$ can be identified with
the space of conformal blocks of $N$ legs of spin $1/2$ and one leg
of spin $j$, and its dimension can be computed using the fusion
rules of the theory.From the previous identification we see that
the centralizer of the quantum group is precisely realized on
the space of conformal blocks , while they remain
invariant under the action of the quantum group. This explains
in a neat way the meaning of hidden quantum symmetries when talking
about the role of quantum groups in RCFT's.

In more physical terms
the centralizer has the structure of a braid group when acting on
the space of conformal blocks. In the case of a WZW model based
on the group $SU(2)$ it was shown by Kanie and Tsuchiya that the
braid group representation that one obtains from the conformal blocks
with spin $1/2$ primary fields at the external legs gives a representation
of the Hecke algebra $H_N(q)$
or more precisely of the Temperley-Lieb-Jones algebra.
On the other hand Kirillov and Reshetikhin
showed that the centralizer of the quantum group $U_q (Sl(2))$
in the tensor product of spin $1/2$ representations was also a
Hecke algebra $H_N(q)$ [KR].
This connection was again part of the phenomenology
that we mention before, however this time one could see that the missing
objects to complete the link with quantum groups were the q-group
spaces $V_{1/2}$ and $V_j$ of equation(5). These spaces should
not be confused with the Verma modules ${\cal{H}}_j$ of the
Kac-Moody algebra. If this were the case then the q-group
$U_q(Sl(2))$ would already be contained in the Kac-Moody algebra
$\widehat{SU(2)}$.
This suggest that the full symmetry algebra should be something like
a tensor product $\widehat{SU(2)} \otimes U_q(Sl(2))$. For a general
conformal field theory the symmetry would be ${\cal{A}}_L \otimes
Q_L$ for the holomorphic degrees of freedom and
${\cal{A}}_R \otimes Q_R$ for the antiholomorphic ones. $Q_L $ and
$Q_R$ denote the quantum groups associated to the chiral algebras
${\cal{A}}_L$ and ${\cal{A}}_R$ respectively. Later on we shall
discuss in more detail the structure of the tensor products
${\cal{A}}_L \otimes Q_L$ which turns out to be a bit more subtle
than just a direct tensor product.
The key idea in this respect is that the q-group spaces $V_j$
contain fields or states that cannot be obtained from the primary
fields by the action of an operator in the chiral algebra
but by that of a genuine quantum group operator.
We should then distinguish between two kind of descendants,
those of the chiral algebra ${\cal{A}}_L$ and those of the quantum group
$Q_L$.

Having settled these conceptual matters the problem was then
to give an "explicit" construction of these quantum group fields
which could not be conformal (either primary or descendants)in the
usual sense. This technical problem was overcome by the use
of the Feigin-Fuchs or Coulomb gas formalism which has proved to
be extremely important in the study of conformal field theories [FF].

\subsection*{The Coulomb Gas formalism of CFT's}

\noindent
A Coulomb gas is nothing but a free field realization of a conformal
field theory. The first known CFT's admiting a Coulomb gas
version are the BPZ $c <1$ theories [BPZ], which were studied by
Dotsenko and Fateev using only one free scalar field [DF]. The CFT's
based on the $W_n$ algebras [Z] require the introduction of n-1
scalar fields [FZ,FL],
while the WZW models need both scalar fields and $\beta-\gamma$
systems [Wa,FFr].
An open problem is to know wether any conformal field theory admits
a free field realization. The fact that a highly interacting field
theory allows a free field description is on the other hand a mistery
with very deep consequences. Some of them are the ability of a systematic
computation of the conformal blocks, the characterization of the
null vectors, a straightforward derivation of the Kac's formula,
the construction of a BRST formulation $ \grave{a} \; la$
Felder [Fe], and , as we shall show
next, the construction of quantum groups.

The basic feature of a Coulomb gas is that the chiral algebra generators
can be constructed entirely in terms of a set of free fields which
have very simple operator product expansions. Moreover the primary
fields, which are essentially vertex operators as in string theory,
also satisfy simple OPE's. All this reduces enormously the
task of computating correlators which are given by integrals
of products of monomials. If we are studying the $c < 1$ or a
WZW model we know that these integrals come from the solution of
hypergeometric type equations. However in the Coulomb gas
these integrals come from the screening charge that one has
to introduce in order to balance the background charge.
A screening charge is an integral $ \int dt J_a(t) $ which has
the property of commuting with all the operators of the chiral
algebra $O_n$ up to a total derivative:

$$
[\; O_n \;, \int dt J_a(t) \; ] = \int dt \frac{ \partial}{ \partial t}
( X_n)
\eqno(6)
$$

\noindent
If there are not boundary contributions to (6) then the screening
charges fully commute with the chiral algebra.

Another ingredient of this construction is the
use of Fock spaces of free fields instead of Verma modules .
The later are recovered as cohomological classes associated
to a BRST charge constructed using screening operators [Fe]. The philosophy
adopted in references [GS1-2] was to use these Fock spaces to enlarge,
rather than to restrict, the number of fields in the theory in
order to give room to q-groups. We could say in this sense
that quantum groups are hidden in a RCFT in much the same manner
that the corresponding free field version is hidden, which lead
us in the long run to ask why the solution to the null
vector decoupling equations of a RCFT admit an integral representation.
This is a highly non trivial fact and a mistery which
certainly needs more study.

\subsection*{Field theory formulation of quantum groups}

\noindent
The starting point of the work of reference [GS1] is the definition
of a special kind of screened vertex operators which form the basis
for the representation spaces of quantum groups. They are
defined integrating a collection of screening vertex operators
{ $J_a$}  around
a primary field $\phi_{\alpha}$:

$$
e^{\alpha}_{a_1,\dots,a_n}( z_P , z_{\infty})
= \int_{C_1} dt_1 \;J_{a_1}(t_1)
\cdots \;\;  \int_{C_n} dt_n \;J_{a_n}(t_n)
\;\; \phi_{\alpha}(z_P)
\eqno(7)
$$

\noindent
In the choice of the
integration contours $C_1,\cdots,C_n$ we have
taken into account the
fact that there are branch cuts in the integrand coming
from the OPE's between the screening operators
with the primary field as well as among the
screening themselves.They are choosen as a nested set
of Hankel's contours, similar to the
ones that appear in the definition of the $\Gamma$
function, i.e. they go from infinity to infinity encercling the
point where the primary field is inserted.
Strictely speaking the operators $e^{\alpha}_{a_1,
\cdots,a_n}$  depend on two points, one is
the point  $z_P$ where the primary field is located and the other
is the point at infinity $z_{\infty}$, which plays the role of a base point.

In this construction the primary field itself
$\phi_{\alpha}$ , is identified with the highest weight vector of a
q-group space $ V_{\alpha}$.
The remaining vertices obtained integrating screening
operators are identified with q-group descendants. This leads naturally
to the interpretation of the screening charges as quantum lowering
operators : $F_a = \int_C dt J_a (t)$.
This interpretation was already anticipated in references [Sa,BMP].
In the $c <1$ theory
there are two screening operators $J_+$ and $J_-$ hence we have two
lowering operators $F_+$ and $F_-$ [GS2]. In the Wakimoto construction
of $\widehat{SU(n)}$ there are n-1 screenings
operators which then lead to n-1 lowering operators each one
associated to a positive simple root [RRR]. Under this interpretation each
screening charge $J_a$
give rise to a quantum group $U_{q_a}(Sl(2))$, where
the deformation parameter $q_a$ is given by the braiding factor among
two screenings of the same type.
If there are various screening operators they will define in general
larger q-groups whose defining relations
would follow from their braiding
properties. For example for a WZW theory
the braiding between screening operators:
satisfies:

$$
J_a(z) \; J_b(w) = q^{C_{a b}} \; J_b(w) \; J_a(z)
\eqno(8)
$$

\noindent
where $C_{a b}$ is the symmetrized Cartan matrix and q is
given by eq.(4).

Equation (8) reflects the non-local nature of the screening
operators and in that sense the quantum deformation parameter
q acquires a conformal field theory meaning.

{}From this equation one can also derive the quantum deformation of
the Serre relations satisfied by the lowering operators $F_a$
[BMP,RRR].

After applying a sufficient number of screening operators to a primary
field one eventually gets zero or a null vector.
To understand why this happens we have to recall that the q-vertex
operators do depend in general on the point at infinity  $z_{\infty}$
and the point $z_P$ where the primary field is inserted, however
under certain circunstances the branch cut effectively disappears
and the contour integrals shrink to the neighbour of the point $z_P$.
This property holds
for primary fields within the Kac's formula and it is extensively
used in the BRST construction. In our q-group interpretation
this means that the primary field give rise to a finite
dimensional representation. In other words, a null vector in the CFT
is at the same time a null vector for the q-group.
This is the
core of the connection between quantum groups and RCFT's.

It would be of some interest to find a more intrinsic definition
of the q-vertex operators (7) , i.e. a definition independent of
the use of the Coulomb gas formalism. A possibility is to
consider these vertices as the solutions
of the
null vector decoupling equations.The primary fields themselves would
of course be a solution, the one with conformal properties, while
the others will span the rest of a representation space.
The order of the equation ,which is nm for the primary field
$\phi_{n m}$ for a $c < 1$ theory, would then give the dimension of
the associated representation space of the quantum group.

We conclude that screening charges are lowering operators, similarly
the Cartan operators are identified with the Coulomb charges. These
two sets of operators form the Borel subalgebra which is traditionally
associated with the lower triangular matrices. To complete the
RCFT picture of q-groups one needs a definition of
the raising operators $E_a$. These cannot be represented as integrals
of screening operators with opposite charge because they do not
have the correct conformal properties. What one really needs
is an operator that would destroy integrals of screening operators
and this can be achieved with the action of the chiral algebra.
Take for example the Virasoro operator $L_{-1}$ , acting on a
q-vertex operator it integrates each of the screening charges
leaving a boundary term as can be seen in eq.(6). Explicit computations
show that eq.(6) implies the q-group relation:

$$
[ O_n , \int dt J_a(t) ] = \int dt \frac{ \partial}{ \partial t}
( X_n) \Rightarrow
[E_a \;, F_b ]= \delta_{a,b} \frac{K_a - K^{-1}_a}{q_a - q^{-1}_a}
\eqno(9)
$$

In this sense the q-group generators $E_a$ are contained in the chiral
algebra generators $O_n$, which shows that the relation between
${\cal{A}}_L$ and $Q_L$ is more intricated than expected.
{}From equation (9) we also see that the shrinking condition for
the contours which define a q-vertex operator is equivalent
to the statement that this q-vertex operator is a highest
weight vector of the quantum group:

$$
\frac{\partial}{\partial z_{\infty}} \; e^{\alpha}_{a_1,\cdots,a_n}
= 0 \Rightarrow E_a \; e^{\alpha}_{a_1,\cdots,a_n}=0 \; \forall a
\eqno(10)
$$

It is also
curious to observe some parallelism between this construction
and the construction of the quantum double by Drinfeld [GS1,BL1].

Having defined the q-group spaces and the q-group generators,
it is easy to obtain the comultiplication, the R-matrix, the
3j-symbols, the 6j-symbols , that is all the ingredients of
quantum groups in a field theoretical framework.
These ideas and techniques, which were
first applied to the $c < 1$ theories, have been extended successfully
to WZW models [RRR], $ W_n$ algebras [C] , N=1 and 2 superconformal
field theories [Jz] and a WZW model based on the supergroup
$Gl(1,1)$ [RSa] .
Moreover the whole
construction has a geometrical and topological flavor [GS3].
Everything follows from contour manipulations
involving elementary complex analysis and q-combinatorics.
Group theory when quantum deformed becomes a topology of contours
(for a more elaborated version of this idea see [FeW,SV]).

\subsection*{Towers of algebras}

\noindent
Until now we have proceed to unravel the quantum group structure
of RCFT's from the integral representation of the decoupling equations,
we can refer to this line of thought as the analytic approach.
A complementary point of view partially inspirated in Pasquier's
ethiology of IRF-models [P] consist in starting the construction
with the fusion rules and from that piece of information to
obtain the duality ( braiding and fusion ) matrices in the very
same way one associates in the Jone's fundamental construction
a Temperley-Lieb-Jones algebra to a given graph [GHJ]. In this algebraic
approach the RCFT defines a model of the Bratelli diagram in terms
of the fusion rules[GS4]. Recent progress by Ocneanu [Oc] indicates
,as it should be expected, that associated with the RCFT tower
of algebras it is possible to define in a unique way a quantum
group structure. For a different version of the algebraic approach
based on operator algebras see [FrK]. We believe that it would be
fruitful both from a mathematical and a physical point of view
to get a deeper understanding of the interplay
between these two apparently different mathematical approaches,
namely the monodromy of differential equations with their related
Riemann-Hilbert problem and the theory of subfactors.

\subsection*{Cannonical quantization approach to CFT's}

\noindent
The discussions above about the role of quantum groups
in CFT's have been done in the general framework of the bootstrap
program of BPZ, where actions for the local fields are not strictely
needed. In doing so we have omitted another important approach
to the role of quantum groups in quantum field theories, namely
the one based on the use of local actions or hamiltonians which are
quantized in a cannonical way. The traditional methods were applied to
Liouville theory which was shown to exhibit quantum group attributes
in a time
when quantum groups were not known [GN,FT].In fact
in reference [FT] first appeared the defining relations
of the quantum Lie group $SL_q$, not to be confused with the
quantum Lie algebra $sl_q$ which is the same as the
quantized universal
enveloping algebra $U_q(Sl(2))$ .

After quantum groups came to fashion this approach was renewed
again in connection with Liouville theory  [Bb,Ge,ST]
, Toda field theory  [BG,HM] and the Wess-Zumino-Witten model
[AS,Fa,Ga,AF,AFS,AFSV] . The use of a different language
has made a bit difficult to stablish the connections with
the, let's say, bootstrap approach although they certainly do
exist. Particularly interesting are the lattice formulations
of the Liouville and the WZW theories which may perhaps allow a Bethe
ansantz analysis analogous to the one applied in the
study of magnetic chains.We refer to the lectures of
prof.L.D.Fadeev for a discussion of these matters.

\subsection*{Integrable quantum field theories}

\noindent
So far we have considered applications of the quantum groups to
conformal field theories but their range of applicability includes
also massive integrable 2D field theories. An archetype
 of the later is the well known sine-Gordon theory. In references
[RS,BL2] it was shown that there exist a hidden quantum symmetry
which governs the scattering
of the solitons, antisolitons and bound states of the model,
which explains in particular the S-matrix obtained in [ZZ].
This symmetry is affine and can also be given a field theory
realization in terms of non-local conserved charges
[BL3,FL,BFL].

\subsection*{q-RCFT's: New hidden symmetries ?.Elliptic
quantum groups}

We would like now to discuss briefly some new directions
of research.
As we have seen, a particular interesting
class of RCFT's are the WZW models based on Kac-Moody
algebras. A  centrally
extended affine Lie algebras can also be deformed
in a very non trivial way. These q-RCFT's preserve much of the structure
of the ordinary RCFT's and can be interpreted as some kind of massive
field theories. In reference [FR] Frenkel and Reshetikhin have found
a q-deformation of the Kniniznik-Zamolodchikov equation
which governs the correlators of these q-WZW model. This is a finite
difference equation whose solutions are q-hypergeometric
functions. The connection matrices, or in CFT language, the braiding
matrices of this "q-conformal blocks" turns out to be given by the
elliptic solution to the Yang-Baxter equation of the RSOS model of
Andrews, Baxter and Forrester [ABF]. These authors also pose the question
of the possible existence of a hidden quantum symmetry which would
underlie the elliptic Boltzmann weights, much in the same way that
quantum groups underlie the trigonometric solutions.
It has also been suggested
as a candidate for this "elliptic quantum symmetry"
the Sklyanin algebra which is intimately related to the Baxter's
eight vertex model [Sk].

The solution to these questions is not known to us. We would
like however to propose a way to attack the problem
which consist in following the same steps that were pursued
in the unravelling of quantum groups in RCFT's, namely :
find a free field realization of the q-Kac-Moody algebra,
then look for the q-screening operators and finally define
in terms of q-integrals the "elliptic version" of
quantum groups.
The first two steps of this program have
already been achieved in references [FJ,M,Sh,KQS,ABG] where various free field
versions of the q deformed Kac-Moody algebra have been given.
In [Sh] it is also constructed a q-screening operator whose
q-integral $\grave{a} \; la$ Jackson commutes with the generators of
q-Kac-Moody.
Therefore one has in principle
all the ingredients to unravel a new hidden
quantum symmetry.

\subsection*{Spin chains at roots of unity}

To finish this brief history we would like to make some comments
in connection to recent progress by De Concini and Kac [DK]
in the study of the representation theory of quantum groups at roots
of unity (see also [DKP].
As we have already mentioned the quantum deformation
parameter of the quantum group associated
to a RCFT is a root of unity , for example
$ q = e^{ \pi i / k + 2}$ for a SU(2) WZW model.
This implies that strictely speaking the quantum group associated
to this RCFT should be $U_q( SU(2))$ moded out by the central
Hopf which appears when $q$ is a root of unity [McSo].

Another phenomena that appears for quantum groups at roots
of unity is the existence of non restricted representations, i.e.
representations which transform non trivially under the central
Hopf subalgebra. In a series of papers
[GRS,GS5,BGS,CGS] we have started the
study of a family of these representations called
semicyclic and
nilpotents, which have the property that their tensor
product is decomposable [GS5,A] so that one
can define 3j and 6j symbols. An important question is wether
this new class of braiding and fusion matrices may become
the duality data of a new class of decoupling equations
defining a new hierarchy of conformal field theories.

\subsection*{Acknowledgements}

We would like to thanks Prof.M.L.Ge for his kind invitation to participate
in the XXI DGMTP Conference at Tianjin.

\end{document}